\documentclass{article}
\usepackage{spconf,amsmath,graphicx,mathtools,amsfonts}
\usepackage{subfigure}
\usepackage{multirow}
\usepackage{booktabs}
\usepackage[colorlinks]{hyperref}

\newcommand{\tabincell}[2]{ \begin{tabular}{@{}#1@{}}#2\end{tabular} }

\title{MEMORY-EFFICIENT LEARNED IMAGE COMPRESSION WITH PRUNED HYPERPRIOR MODULE}

\name{Ao Luo$^1$, Heming Sun$^{2,3}$, Jinming Liu$^1$, Jiro Katto$^1$}
\address{$^1$Department of Computer Science and Communication Engineering, Waseda University, Tokyo, Japan\\
$^2$Waseda Research Institute for Science and Engineering, Tokyo, Japan\\
$^3$JST, PRESTO, 4-1-8 Honcho, Kawaguchi, Saitama, Japan}

\begin{document}
%
\maketitle
\begin{abstract}

Learned Image Compression (LIC) gradually became more and more famous in these years. The hyperprior-module-based LIC models have achieved remarkable rate-distortion performance.
However, the memory cost of these LIC models is too large to actually apply them to various devices, especially to portable or edge devices. The parameter scale is directly linked with memory cost. In our research, we found the hyperprior module is not only highly over-parameterized, but also its latent representation contains redundant information. Therefore, we propose a novel pruning method named ERHP in this paper to efficiently reduce the memory cost of hyperprior module, while improving the network performance. The experiments show our method is effective, reducing at least 22.6\% parameters in the whole model while achieving better rate-distortion performance.

\end{abstract}
\section{Introduction}
\label{sec:intro}

Image compression is one of the important part for various applications. The conventional compression standards, such as JPEG \cite{jpg}, JPEG 2000 \cite{jpg2000}, BPG (Better Portable Graphics) \cite{bpg} and VVC (Versatile Video Coding) \cite{vvc} mainly use linear transform with hand-designed codes to compress the images. In recent several years, deep-learning based image compression (Learned Image Compression, LIC) methods take use of neural network, which has non-linear activation, to compress the images. These methods gradually outperform the classic ones.

One famous LIC method is Hyperprior \cite{hyperprior}, which utilizes a hyperprior module to capture the spatial redundancy among neighboring elements. The whole model consists of two parts: main path $g_a$, $g_s$ and hyper path $h_a$, $h_s$, both of which are pairs of encoder $*_a$ and decoder $*_s$, as shown in Fig. \ref{fig:hyperprior}.
The main path receives input image, generating its latent representation $y$, which is assumed to obey arbitrary zero-mean Gaussian distribution. Then the hyper encoder uses $y$ to calculate the side information $z$, which helps the hyper decoder to generate scale of $y$. With scale information, $y$ is rescaled to standard normal distribution and transmitted with $z$ together to decoder part. Regarding to scale generated by $z$, $y$ is generated from transmitted information and is given to main decoder to reconstruct the image.
With the help of hyper path, Hyperprior model achieved dramatic progress, exceeding conventional methods. Regarding to its good performance, hyperprior module became an important part in latter methods, such as the methods improving performance \cite{joint, sun, gmm}, the ones making LIC models more appliable \cite{slimmable, checkerboard}.

However, one problem for LIC is that the requirements for memory are much larger than the conventional methods. With the development of mobile internet, more and more people tend to display and store images on portable or edge devices, whose memory is not sufficient for an image compression algorithm, such as mobile phones, advertising screens and so on. Therefore, it is difficult for LIC methods to widespread in practice.

The memory cost is directly linked with parameter scale. There are some former works focusing on lightweight image compression models with much lower FLOPs and parameter scale. \cite{acmmm} proposed several lightweight components, which decreased FLOPs and memory cost. \cite{googlelight} implements group Lasso loss to prune convolution layers' channels in decoding part, by which the researchers obtained lightweight models. However, the rate-distortion performance of these works dropped distinctly. In the meanwhile, \cite{googlelight} also found that their pruning method has little impact on hyper path.

In this paper, based on ResRep \cite{resrep}, we propose a pruning method called ERHP (Enhanced Resrep on Hyper Path in learned image compression) to prune LIC models. We finetune the pruned network to recover its performance, since the image compression task requires higher quality than image classification task on which ResRep is proposed. In this way, our method reduces parameter scale distinctively while even improving the performance. Our experiments on Hyperprior model\cite{hyperprior} and Cheng\cite{gmm} show the efficiency of that our method.

To summarize, our contributions are listed as below:
\begin{itemize}
    \item We confirmed that hyper path is severely over-paramet-erized, which can be pruned to reduce memory consumption and redundancy.
    \item We propose a ERHP, which adapts ResRep pruning method to the LIC task by implementing PixelShuffle \cite{pixel} layer and deconv layer, and prunes the LIC models efficiently.
    \item The experiments on Hyperprior model\cite{hyperprior} and Cheng\cite{gmm} show our method achieves much lower parameter scale and even improves the performance of the pruned model. In this way, the LIC models are more applicable for edge devices and perform better than the former models.
\end{itemize}

The following parts of this paper are arranged as below. In Sec. \ref{sec:rel}, we introduce the former methods directly linked with our work. After that, our proposed ERHP is introduced in Sec. \ref{sec:method}. Then, the experiment results are shown in Sec. \ref{sec:exp}. Finally, we summarize our work in Sec. \ref{sec:con}.

\section{Related Work}\label{sec:rel}

There have been plenty of works on lightweight models and cropping parameters. There are lots of former works taking use of Lasso loss penalty, such as \cite{sunjian, realtime}. The Lasso loss penalty calculates $l1$ regularization of weights ($\|W\|_1$ in Eq. \ref{eq:lasso}) and add this penalty term to loss function with a coefficient $\beta$. In training step, the penalty term suppresses weights to zero, and the other terms in loss function amplify the weights. With the Lasso loss, part of weights are reduced to zero, while keeping others valid enough.

\begin{equation}
    \label{eq:lasso}
    \mathcal{L}_{prune1} = \mathcal{L}_{quality} + \beta\cdot\|W\|_{1}
\end{equation}

Johnston et. al \cite{googlelight} applied to LIC the Group Lasso loss \cite{grouplasso}, which groups the weights in a convolution kernel (3x3 or 5x5 and so on) together by $l2$ norm, as defined in Eq. \ref{eq:group},

\begin{equation}
    \label{eq:group}
    \mathcal{L}_{prune2}=\mathcal{L}_{quality}+\beta\cdot\sum\|w_i\|_2
\end{equation}
where $w_i$ is the $i$-th kernel in the network. This work cropped a lot of parameters, but the performance of output models dropped too.

ResRep \cite{resrep} splits the weights for remembering and pruning. It add an 1x1 convolution layer behind each of the convolution layers to be pruned. This 1x1 convolution layer, called compactor, has the same input and output channel. The weights of compactor are initialized as identity matrices, whose output is the same as its input. The Group Lasso loss is only applied to the compactor, while the normal loss is applied to all the components. When finishing the pruning, ResRep combines each pair of original convolution layer and compactor together. In this way, the network retains good performance and is pruned efficiently. However, ResRep just provided components for normal CNN, which cannot satisfy LIC models.

\section{Proposed Method}\label{sec:method}

\subsection{Preliminary Analysis}

In this section, we illustrate our analysis on hyper path, introducing the differences between main path and hyper path.

\begin{figure}[tbp]
    \centering
    \includegraphics[width=.45\textwidth]{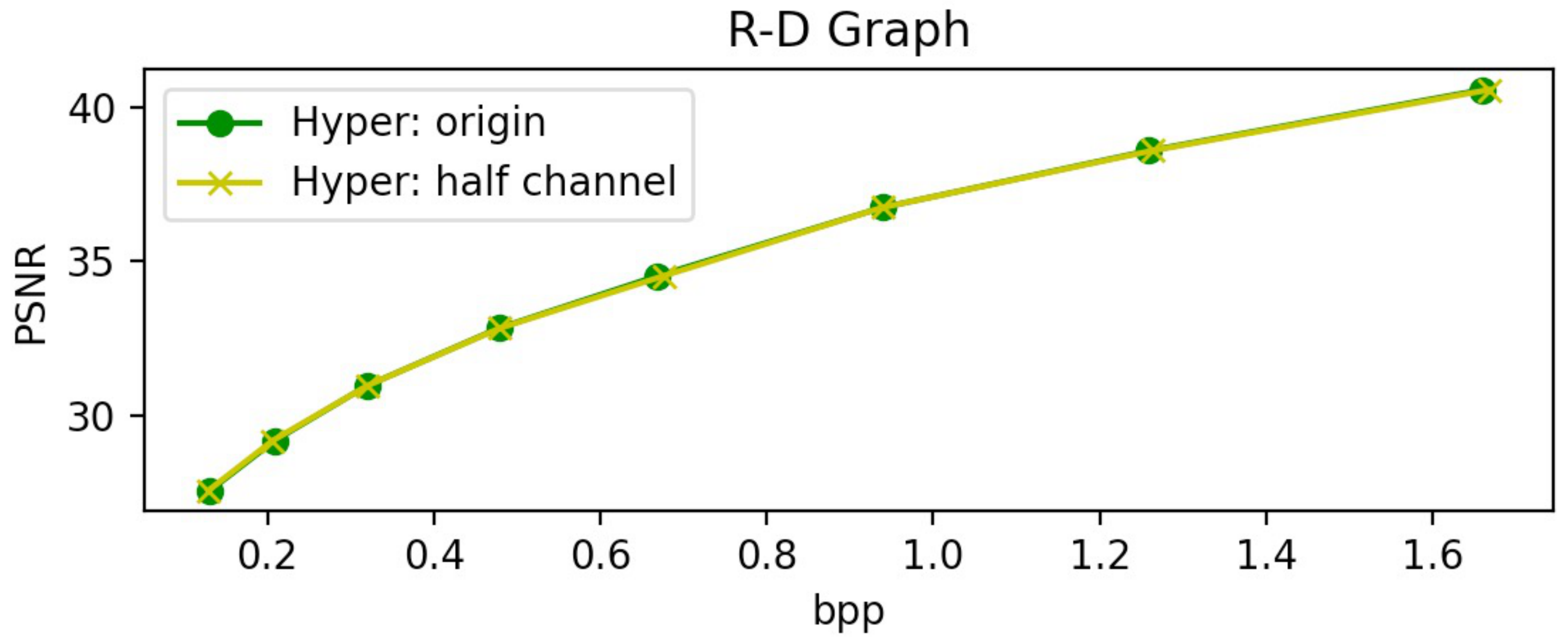}
	\caption{Comparison between original and half-channel-pruned hyperprior model.}
    \label{fig:hyper}
\end{figure}

The parameter scale determines the upper bound of information volume can be obtained by the network. In our experiments, we compared the bit rate and parameter scale between main path ($y$) and hyper path ($z$). In hyperprior model \cite{hyperprior}, the parameter ratio of hyper path is approximately 40\%, while the bit-rate ratio of $z$ is only 1.7\%. When inferring, the feature map of $z$ is much smaller than that of $y$, which means the information $z$ carries is distinctively lower than $y$. This phenomenon reveals the hyper path is heavily over-parameterized. We confirmed this conclusion in our experiments. As shown in Fig. \ref{fig:hyper}, even cropped half of channels in hyper path, the models still achieves almost the same performance as original models.

On the other hand, \cite{hyperprior} mentioned the relationship between channel numbers and rate-distortion performance. With more channels in the network, more details are extracted and obtained, finally lead to better performance.
However, they set the same channels for main path and hyper path in their experiments, which ignored the difference between them.
It is true that, with more channels in main path, the network extracts more details from the original image, which significantly improve the reconstructed image. However, hyper path generates the scale information to rescale $y$ into an approximate range, which requires not so much details. In our experiments, more channels in hyper path even cause performance decline because of information redundancy.

\subsection{Problem Formulation}

We formulate the LIC task and introduce ResRep to better illustrate our proposed method.

\subsubsection{Learned Image Compression with Hyperprior}

\begin{figure}[tbp]
    \centering
    \includegraphics[width=.40\textwidth]{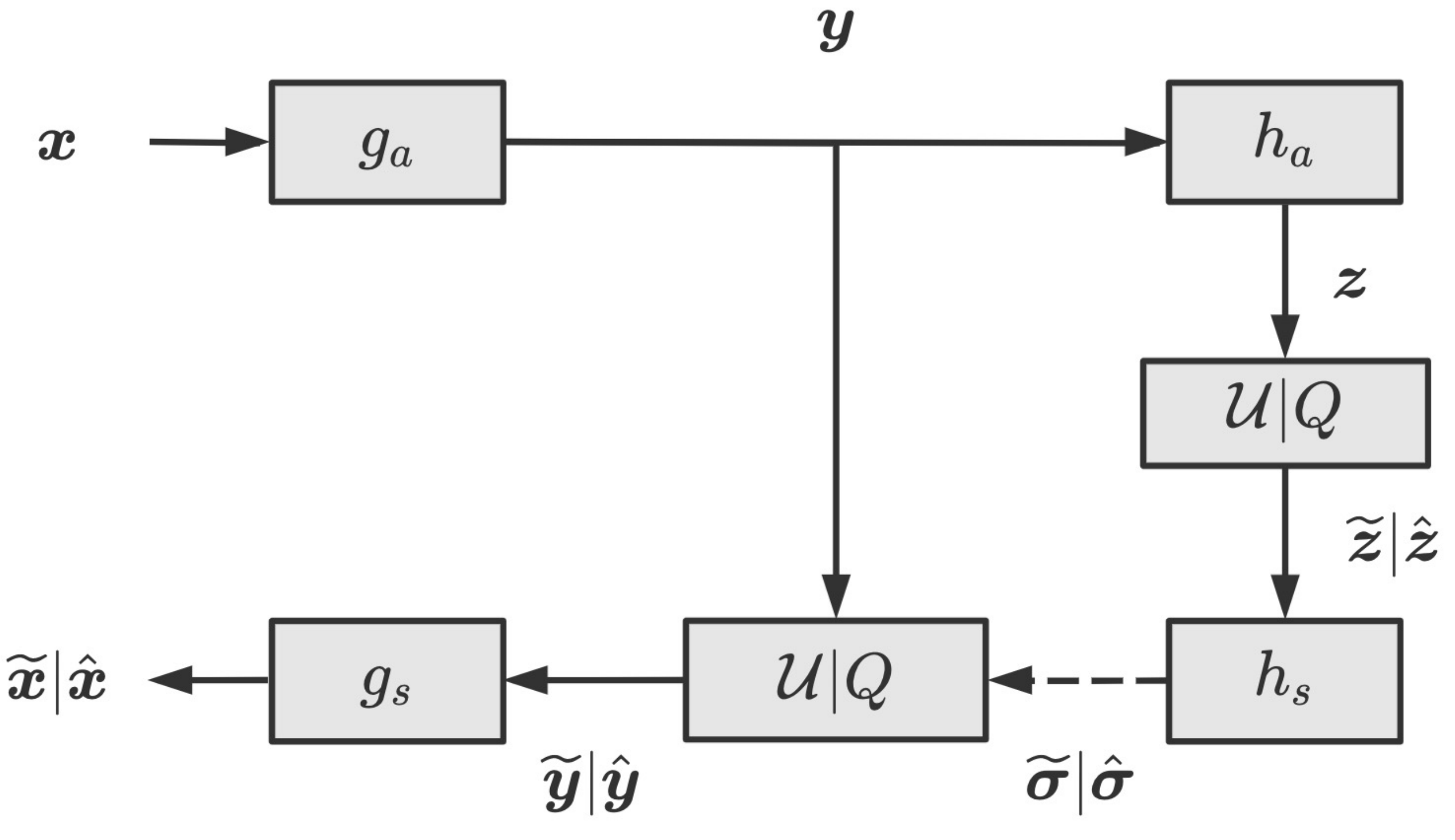}
	\caption{Sketch of hyperprior model.}
    \label{fig:hyperprior}
\end{figure}

As mentioned before, the hyperprior model \cite{hyperprior} is shown in Fig. \ref{fig:hyperprior}. $g_a$, $g_s$, $h_a$, $h_s$ are nonlinear neural networks. $x$ is the input image, $y=g_a(x)$ and $z=h_a(y)$ are latent representation and hyper latent, respectively. $\hat{y}=Q(y)$ and $\hat{z}=Q(z)$ are quantized $y$ and $z$. $\hat{z}$ is taken as side information for generating the scale parameter $\hat\sigma$ for the entropy model of latent $\hat{y}$. $\hat{x}=g_s(\hat{y})$ is the reconstructed image. In training step, the quantization operation is applied by adding uniform noise, which produces differentiable variables $\widetilde{y}$, $\widetilde{z}$, $\widetilde{x}$ and $\widetilde\sigma$. For simplicity, we represent $\widetilde{x}|\hat{x}$, $\widetilde{y}|\hat{y}$, $\widetilde{z}|\hat{z}$ and $\widetilde\sigma|\hat\sigma$ as $\hat{x}$, $\hat{y}$, $\hat{z}$ and $\hat\sigma$. The loss function can be written as the trade-off of rate $R$ and distortion $D$:

\begin{equation}
\begin{aligned}
    \mathcal{L}_{LIC}=R+\lambda D=&\mathbb{E}_{x\sim p_x}\left[-log_2p_{\hat{y}|\hat{z}}(\hat{y}|\hat{z})-log_2p_{\hat{z}}(\hat{z})\right]\\
    &+\lambda \cdot \mathbb{E}_{x\sim p_x}\left[d(x,\hat{x})\right]
    \label{eq:loss}
\end{aligned}
\end{equation}
where the bit rate of $\hat{y}$ and $\hat{z}$ is evaluated by entropy, $\lambda$ controls the trade-off of rate and distortion, $p_*$ is probability of $*$, and $d(x,\hat{x})$ is the distortion, which is MSE in our work.

\subsubsection{ResRep Pruning Method}

The weight and bias of the $i$-th convolution layer can be shown as $\boldsymbol{W}_{i}\in \mathbb{R}^{C_{i+1}\times C_{i}\times ks\times ks}$ and $\boldsymbol{b}_i\in \mathbb{R}^{C_{i+1}}$, where $C_{i+1}$ and $C_{i}$ is the input channel number of layer $i$ and $i-1$, and $ks$ is the kernel size of this layer. The weight of compactor of the $i$-th convolution layer is $\boldsymbol{R}_{i}\in \mathbb{R}^{C_{i+1}\times C_{i+1}\times1\times1}$, which is initialized as an identity matrix. After pruning, the $\boldsymbol{R}_{i}$ is cropped as $\boldsymbol{R}_i^{'}\in \mathbb{R}^{C_{i+1}^{'}\times C_{i+1}\times1\times1}$, where $C_{i+1}^{'}$ is the pruned channel number, satisfying $C_{i+1}^{'}\leq C_{i+1}$. Then ResRep combines the normal convolution layer and compactor layer, as shown in Eqs. \ref{eq:comb} and \ref{eq:trans}:

\begin{equation}
\label{eq:comb}
\begin{aligned}
    \boldsymbol{O} &= \boldsymbol{I}\otimes \boldsymbol{W}_i^{'}+B(\boldsymbol{b}_i^{'})\\
    &= (\boldsymbol{I}\otimes \boldsymbol{W}_i+B(\boldsymbol{b}_i))\otimes \boldsymbol{R}_i^{'}\\
    &= \boldsymbol{I}\otimes \boldsymbol{W}_i^{'}\otimes \boldsymbol{R}_i^{'}+B(\boldsymbol{b}_i)\otimes\boldsymbol{R}_i^{'}
\end{aligned}
\end{equation}

\begin{equation}
\label{eq:trans}
    \boldsymbol{W}_i^{'}=T(T(\boldsymbol{W}_i)\otimes \boldsymbol{R}_i^{'})
\end{equation}
\begin{equation}
\label{eq:transbias}
    b_{i;j}^{'}=\boldsymbol{b}_i\cdot R^{'}_{i;j,:,:,:}, \forall 1\leq j \leq C_{i+1}^{'}
\end{equation}
where $O$ and $I$ are the output and input of the network, $\cdot$ is element-wise multiply, $\otimes$ is the convolution operation, $B(*)$ is the duplication for bias, and $T(*)$ is transposition.

\subsection{Enhanced ResRep on Hyper Path in Learned Image Compression}

The ResRep pruning method only implemented compactor for standard convolution layer and convolution layers with batch normalization, which cannot satisfy LIC models. In this paper, we propose an ERHP (Enhanced Resrep on Hyper Path in learned image compression), which implements compactors for PixelShuffle \cite{pixel} layer and deconvolution layer for adaptation to LIC models.

The PixelShuffle component expands feature map by $\alpha$ times, and arranges neighbour $\alpha^2$ channels into one plain. For example, the shape of input for PixelShuffle operation is $\alpha^2 C_{i+1}\times H\times W$, where the output should be $C_{i+1}\times \alpha H\times \alpha W$. Therefore, every $\alpha^2$ channels in convolution layer of PixelShuffle generate a small patch, and should be reserved or removed together. We put the compactor after the shuffling operation, instead of after the convolution layer directly, as shown in Eq. \ref{eq:pixsfl}:

\begin{equation}
\label{eq:pixsfl}
\begin{aligned}
    \boldsymbol{O}&=PS(\boldsymbol{I}\otimes \boldsymbol{W}_i^{ps} + B(\boldsymbol{b}_i^{ps}))\otimes R_i^{'}\\
\end{aligned}
\end{equation}
where $PS(*)$ is the PixelShuffle operation, and the weight of convolution layer is $\boldsymbol{W}_i^{ps} \in \mathbb{R}^{\alpha^2C_{i+1}\times C_{i}\times ks\times ks}$.

The combination of convolution layer and compactor is
\begin{equation}
\begin{aligned}
    \boldsymbol{W}_{i;k::\alpha^2,:,:,:}^{ps'}&=T(T(\boldsymbol{W}_{i;k::\alpha^2,:,:,:}^{ps})\otimes R_i^{'}), \forall1\leq k\leq \alpha^2\\
\end{aligned}
\end{equation}
\begin{equation}
\begin{aligned}
    b_{i;k\times\alpha^2+j}^{'}=b_{i;k::\alpha^2}\cdot R_{i;j,:,:,:}^{'}, \forall &1\leq k \leq\alpha^2,\\
    &1\leq j\leq C_{i+1}^{'}
\end{aligned}
\end{equation}
where the $k::\alpha^2$ means that, starting from the $k$-th element, select an element every $\alpha^2$. For example, the $k$-th, $k+\alpha^2$-th, $k+2\alpha^2$-th elements are selected.

The deconvolution layer is utilized in decoder part of hyper path. It upsamples the feature map of $\hat{z}$ from bit stream. The structure of its weight is $W^{de}_i\in \mathbb{R}^{C_{i}\times C_{i+1}\times ks\times ks}$, where the input and output channels are transposed, while the bias keeps the same. Therefore, the combination of its weight and corresponding compactor is written as

\begin{table*}[htbp]
\scriptsize
\centering
\caption{Pruning results of ERHP. Quality ($\lambda$) is the trade-off of rate and distortion, as shown in Eq. \ref{eq:loss}. Corresponding PSNRs are the same because of frozen main path and context model.}
\begin{tabular}{c c c c c c c}
\toprule
\multirow{2}*{\textbf{Quality($\lambda$)}} & \multicolumn{3}{c}{\textbf{Results of Hyperprior Model\cite{hyperprior}}} & \multicolumn{3}{c}{\textbf{Results of Cheng\cite{gmm}}}\\ \cmidrule(r){2-4} \cmidrule(r){5-7}
~ & \tabincell{c}{Origin Performance\\PSNR@BPP} & \tabincell{c}{ERHP Performance\\PSNR@BPP} & \tabincell{c}{Parameter Scale\\pruned/origin} & \tabincell{c}{Origin Performance\\PSNR@BPP} & \tabincell{c}{ERHP Performance\\PSNR@BPP} & \tabincell{c}{Parameter Scale\\pruned/origin}\\ \hline
\toprule
$0.0483$ & 36.706@0.937 & 36.706@\textbf{0.936} & 7.748M/11.582M(33.1\%$\downarrow$) & 36.898@0.823 & 36.898@\textbf{0.818} & 21.867M/28.244M(22.6\%$\downarrow$) \\ \hline
$0.0250$ & 34.501@0.667 & 34.501@\textbf{0.667} & 3.705M/4.969M(25.4\%$\downarrow$) & 35.282@0.603 & 35.282@\textbf{0.601} & 21.576M/28.244M(23.6\%$\downarrow$) \\ \hline
$0.0130$ & 32.823@0.478 & 32.823@\textbf{0.476} & 3.651M/4.969M(26.5\%$\downarrow$) & 33.521@0.433 & 33.521@\textbf{0.429} & 21.327M/28.244M(24.5\%$\downarrow$) \\ \hline
$0.0067$ & 30.962@0.319 & 30.962@\textbf{0.319} & 3.589M/4.969M(27.8\%$\downarrow$) & 31.318@0.292 & 31.318@\textbf{0.290} & 9.700M/12.563M(22.8\%$\downarrow$) \\ \hline
$0.0035$ & 29.192@0.209 & 29.192@\textbf{0.208} & 3.264M/4.969M(34.3\%$\downarrow$) & 29.763@0.199 & 29.763@\textbf{0.197} & 9.663M/12.563M(23.1\%$\downarrow$) \\ \hline
$0.0018$ & 27.578@0.131 & 27.578@\textbf{0.131} & 3.318M/4.969M(33.2\%$\downarrow$) & 28.233@0.130 & 28.233@\textbf{0.127} & 9.526M/12.563M(24.2\%$\downarrow$) \\
\toprule
\end{tabular}
\label{tab:gmm}
\end{table*}

\begin{table}[htbp]
\centering
\small
\caption{Comparisons of different pruning methods on Cheng\cite{gmm}.}
\begin{tabular}{c c c c c}
\toprule
\multirow{2}*{\textbf{Quality($\lambda$)}} & \multirow{2}*{\textbf{PSNR}} & \multicolumn{3}{c}{\textbf{BPP}}\\ \cline{3-5}
~ & ~ & Origin & Manual & ERHP \\
\toprule
$0.0483$ & 36.898 & 0.823 & 0.983 & \textbf{0.818}\\ \hline
$0.0250$ & 35.282 & 0.603 & 0.601 & \textbf{0.601}\\ \hline
$0.0130$ & 33.521 & 0.433 & 0.429 & \textbf{0.429}\\ \hline
$0.0067$ & 31.318 & 0.292 & 0.292 & \textbf{0.290}\\ \hline
$0.0035$ & 29.763 & 0.199 & 0.197 & \textbf{0.197}\\ \hline
$0.0018$ & 28.233 & 0.130 & 0.129 & \textbf{0.127}\\ \hline
\toprule
\end{tabular}
\label{tab:manual}
\end{table}

\begin{equation}
\begin{aligned}
    \boldsymbol{W}^{de'}_i&=\boldsymbol{W}^{de}_i\otimes \boldsymbol{R}^{'}_i
\end{aligned}
\end{equation}
where the converting of bias is the same with Eq. \ref{eq:transbias}.

With the newly implemented two components, we apply ResRep pruning method on hyper path. The whole loss function for pruning is written as
\begin{equation}
\begin{aligned}
    \mathcal{L}
    =&R+\lambda D+\beta\mathcal{L}_{lasso}\\
    =&\mathbb{E}_{x\sim p_x}\left[-log_2p_{\hat{y}|\hat{z}}(\hat{y}|\hat{z})-log_2p_{\hat{z}}(\hat{z})\right]\\
    &+\lambda \cdot \mathbb{E}_{x\sim p_x}\left[d(x,\hat{x})\right]\\
    &+\beta\cdot\sum_{i=1}^L\sum_{j=1}^{C_i}\|\boldsymbol{R}_{i;j,:,:,:}\|_2
\end{aligned}
\end{equation}
where $\lambda$ is the rate-distortion trade-off, $\beta$ is the lasso panelty strength, $\boldsymbol{R}_{i;j,:,:,:}$ is the $j$-th channel of the $i$-th layer to be pruned, and $L$ means the total number of layers to be pruned.

In the original ResRep, the authors directly use the pruned model to classify images. However, the image compression task requires higher latent representation qualities than image classification task, so we finetune the pruned model to recover the original performance.

\section{Experiments}\label{sec:exp}

In this section, we show our implement details and experiment results of ERHP-pruned models on Hyperprior model\cite{hyperprior} and Cheng\cite{gmm}, which achieve distinctively lower memory cost and improve the rate-distortion performance.

\subsection{Implement Details}

We set the lasso strength $\beta$ to 1e-9, which is small enough to keep convolution layers available, while effectively pruning the model. We set the preliminary pruning target to 0.7, which means the initial aim is to prune 70\% parameters in hyper path and is appropriate for most of the models. After pruning, we use \cite{cai} to finetune our model, with learning rate of 1e-4.

\subsection{Experiments of Pruned Models}

We trained our models on OpenImage \cite{openimage} and tested them on Kodak \cite{kodak}. The PSNR ($10log_{10}\frac{255^2}{mse}$) and bit-per-pixel (bpp) are taken as our evaluating metrics.

We treat models as two parts: hyper path to prune; main path and context model (Hyperprior model\cite{hyperprior} does not have context model) to freeze. To be concrete, when pruning, we initialize the model with pretrained models from \cite{cai}, then train the hyper path only, while main path and context model frozen. We apply compactors to the whole hyper path except the last layer in hyper decoder, since the final output channel cannot be changed according to the fixed dimension of $y$.
Table \ref{tab:gmm} shows the general results of ERHP on Hyperprior model and Cheng model. The results prove the efficiency of our ERHP, achieving at least 22.6\% parameter reduction in the whole model.

We did ablations on ERHP and manually pruned models at the same parameter scale with ERHP results but uniform channel number for each layer in hyper path, as shown in Table \ref{tab:manual}. Because the main path and context model are frozen, the PSNR of same-quality models are the same in our experiments. Our method not only solves the over-parameterize problem, but also reduces the redundancy in $z$, improving the performance of the models slightly. All the results of ERHP are better than or equal to the corresponding manually pruned models, which shows the effectiveness of our method.

\section{Conclusions}\label{sec:con}

In this paper, we propose a novel ERHP (Enhanced Resrep on Hyper Path in learned image compression) for reducing the memory cost of LIC models by pruning channels of hyper path. We perform compactors for PixelShuffle and deconvolution layer, which are used in LIC models. The experiments on Hyperprior model and Cheng model show that our method is effective, pruning a large amount of parameters while improving the rate-distortion performance.

\vfill\pagebreak
\bibliographystyle{IEEEbib}
\bibliography{strings,refs}

\begin{thebibliography}{10}

\bibitem{jpg}
Gregory~K Wallace,
\newblock ``The jpeg still picture compression standard,''
\newblock {\em IEEE transactions on consumer electronics}, vol. 38, no. 1, pp.
  xviii--xxxiv, 1992.

\bibitem{jpg2000}
Majid Rabbani,
\newblock ``Jpeg2000: Image compression fundamentals, standards and practice,''
\newblock {\em Journal of Electronic Imaging}, vol. 11, no. 2, pp. 286, 2002.

\bibitem{bpg}
Fabrice Bellard,
\newblock ``Bpg image format,''
\newblock {\em https://bellard.org/bpg}, 2015.

\bibitem{vvc}
Benjamin Bross, Jianle Chen, Shan Liu, and Ye-Kui Wang,
\newblock ``Jvet-s2001 versatile video coding (draft 10),''
\newblock in {\em Joint Video Exploration Team (JVET) of ITU-T SG 16 WP 3 and
  ISO/IEC JTC 1/SC 29/WG 11}, 2020.

\bibitem{hyperprior}
Johannes Ballé, David Minnen, Saurabh Singh, Sung~Jin Hwang, and Nick
  Johnston,
\newblock ``Variational image compression with a scale hyperprior,''
\newblock in {\em International Conference on Learning Representations}, 2018.

\bibitem{joint}
David Minnen, Johannes Ball\'{e}, and George~D Toderici,
\newblock ``Joint autoregressive and hierarchical priors for learned image
  compression,''
\newblock in {\em Advances in Neural Information Processing Systems}, 2018,
  vol.~31.

\bibitem{sun}
Zhao Zan, Chao Liu, Heming Sun, Xiaoyang Zeng, and Yibo Fan,
\newblock ``Learned image compression with separate hyperprior decoders,''
\newblock {\em IEEE Open Journal of Circuits and Systems}, vol. 2, pp.
  627--632, 2021.

\bibitem{gmm}
Zhengxue Cheng, Heming Sun, Masaru Takeuchi, and Jiro Katto,
\newblock ``Learned image compression with discretized gaussian mixture
  likelihoods and attention modules,''
\newblock in {\em Proceedings of the IEEE/CVF Conference on Computer Vision and
  Pattern Recognition}, 2020, pp. 7939--7948.

\bibitem{slimmable}
Fei Yang, Luis Herranz, Yongmei Cheng, and Mikhail~G Mozerov,
\newblock ``Slimmable compressive autoencoders for practical neural image
  compression,''
\newblock in {\em Proceedings of the IEEE/CVF Conference on Computer Vision and
  Pattern Recognition}, 2021, pp. 4998--5007.

\bibitem{checkerboard}
Dailan He, Yaoyan Zheng, Baocheng Sun, Yan Wang, and Hongwei Qin,
\newblock ``Checkerboard context model for efficient learned image
  compression,''
\newblock in {\em Proceedings of the IEEE/CVF Conference on Computer Vision and
  Pattern Recognition}, 2021, pp. 14771--14780.

\bibitem{acmmm}
Zekun Zheng, Xiaodong Wang, Xinye Lin, and Shaohe Lv,
\newblock {\em Get The Best of the Three Worlds: Real-Time Neural Image
  Compression in a Non-GPU Environment}, p. 5400–5409,
\newblock Association for Computing Machinery, New York, NY, USA, 2021.

\bibitem{googlelight}
Nick Johnston, Elad Eban, Ariel Gordon, and Johannes Ballé,
\newblock ``Computationally efficient neural image compression,''
\newblock Tech. {R}ep., Google Research, 2019.

\bibitem{resrep}
Xiaohan Ding, Tianxiang Hao, Jianchao Tan, Ji~Liu, Jungong Han, Yuchen Guo, and
  Guiguang Ding,
\newblock ``Resrep: Lossless cnn pruning via decoupling remembering and
  forgetting,''
\newblock in {\em Proceedings of the IEEE/CVF International Conference on
  Computer Vision}, 2021, pp. 4510--4520.

\bibitem{pixel}
Wenzhe el.~al. Shi,
\newblock ``Real-time single image and video super-resolution using an
  efficient sub-pixel convolutional neural network,''
\newblock in {\em Proceedings of the IEEE conference on computer vision and
  pattern recognition}, 2016, pp. 1874--1883.

\bibitem{sunjian}
Yihui He, Xiangyu Zhang, and Jian Sun,
\newblock ``Channel pruning for accelerating very deep neural networks,''
\newblock in {\em Proceedings of the IEEE international conference on computer
  vision}, 2017, pp. 1389--1397.

\bibitem{realtime}
Zheng et.~al. Zhan,
\newblock ``Achieving on-mobile real-time super-resolution with neural
  architecture and pruning search,''
\newblock in {\em Proceedings of the IEEE/CVF International Conference on
  Computer Vision}, 2021, pp. 4821--4831.

\bibitem{grouplasso}
Ming Yuan and Yi~Lin,
\newblock ``Model selection and estimation in regression with grouped
  variables,''
\newblock {\em Journal of the Royal Statistical Society. Series B, Statistical
  Methodology}, pp. 49--67, 2006.

\bibitem{cai}
Jean B{\'e}gaint, Fabien Racap{\'e}, Simon Feltman, and Akshay Pushparaja,
\newblock ``Compressai: a pytorch library and evaluation platform for
  end-to-end compression research,''
\newblock {\em arXiv preprint arXiv:2011.03029}, 2020.

\bibitem{openimage}
Candice~Schumann et. al.,
\newblock ``A step toward more inclusive people annotations for fairness,''
\newblock in {\em Proceedings of the AAAI/ACM Conference on AI, Ethics, and
  Society (AIES)}, 2021.

\bibitem{kodak}
``The kodak photocd dataset,''
\newblock {\em http://r0k.us/graphics/kodak/.}

\end{thebibliography}

\end{document}